**Generics in science communication: Misaligned interpretations across laypeople, scientists, and large language models**

[This is a penultimate draft of a paper forthcoming in *Public Understanding of Science*.]

Uwe Peters,[1] Andrea Bertazzoli,[2] Jasmine M. DeJesus,[3] Gisela J. van der Velden,[4] and Benjamin Chin-Yee[5]

## Abstract

Scientists often use *generics*, that is, unquantified statements about whole categories of people or phenomena, when communicating research findings (e.g., “statins reduce cardiovascular events”). Large language models (LLMs), such as ChatGPT, frequently adopt the same style when summarizing scientific texts. However, generics can prompt overgeneralizations, especially when they are interpreted differently across audiences. In a study comparing laypeople, scientists, and two leading LLMs (ChatGPT-5 and DeepSeek), we found systematic differences in interpretation of generics. Compared to most scientists, laypeople judged scientific generics as more generalizable and credible, while LLMs rated them even higher. These mismatches highlight significant risks for science communication. Scientists may use generics and incorrectly assume laypeople share their interpretation, while LLMs may systematically overgeneralize scientific findings when summarizing research. Our findings underscore the need for greater attention to language choices in both human and LLM-mediated science communication.

[1] Corresponding author: u.peters@uu.nl. Dept of Philosophy, Utrecht University, Netherlands
[2] Dept of Philosophy, Utrecht University, Netherlands
[3] Dept of Psychology, University of North Carolina at Greensboro, USA
[4] Dept of Biomedical Sciences, Education and Research Centre, UMC, Utrecht, Netherlands
[5] Dept. of Pathology and Laboratory Medicine, Medicine, and Philosophy, Western University, Canada

## 1. Introduction

Science has a vital societal function, providing experts, policymakers, and the public with the knowledge required to make informed decisions (Kitcher, 2011). To fulfil its function, science needs to be communicated to people in ways that accurately capture research findings and avoid misinterpretations (Jamieson et al., 2017).

Prior research in science communication has found that different stakeholder groups, including scientists, policymakers, and members of the public, often diverged in their understanding and interpretation of scientific information (e.g., on genome editing technologies; Calabrese et al., 2021; Robbins et al., 2021). Yet less is known about the specific communicative features that contribute to such interpretive differences (e.g., Venhuizen et al., 2019).

One potential source of misalignment may lie in scientists' use of generalizing language. Studies found that when reporting research findings, scientists frequently used *generics*, which are commonly defined as present tense statements expressing generalizations without quantifiers (e.g., "some," "many"), thereby referring to entire categories rather than numerically specified subsets of people, objects, or phenomena (e.g., DeJesus et al., 2019). For example, "people benefit from a Mediterranean diet" is a generic, whereas "some people benefit from a Mediterranean diet" is not. Similarly, "statins reduce cardiovascular events" is a generic, but "many statins reduce cardiovascular events" is not. While generics suggest that results apply to all or most members of a kind (people, statins, etc.) (Cimpian et al., 2010), corpus analyses found that many scientists used them to describe findings even when they were based on small, unrepresentative samples that did not justify such broad claims (Rad et al., 2018; DeJesus et al., 2019; Peters et al., 2024).

Scientists' use of generics can be problematic. Audiences may overgeneralize findings (e.g., about medical treatments) to populations for which they are not applicable (Chin-Yee, 2023). Overly broad claims can also deprive audiences of reliable information for decision-making and undermine public trust in science, especially when inflated expectations go unmet (Master and Resnik, 2013).

The risks of misunderstanding and overgeneralization are particularly high if science communicators and their audiences differ in how they interpret scientific generics. The likelihood of such interpretive misalignments is increased by the fact that generics are ambiguous (Cohen, 2022). For instance, "people benefit from a Mediterranean diet" can refer to some, many, or all people. Scientific expertise including stronger "epistemic vigilance" – the disposition to critically evaluate information source, content, and context (Sperber et al., 2010) – or shared scientific background knowledge ("common ground") (Clark, 1996) may lead scientists to interpret generics more narrowly than laypeople. If so, scientists may use them even though their audience might understand them differently than intended, potentially leading to miscommunication.

Recent studies comparing novices and experts in a domain outside of science (e.g., video gaming) found evidence of such expertise effects and adjustment failures by experts (Coon et al., 2021). However, potential differences in generics interpretation between scientists and laypeople remain unexplored (Haigh et al., 2020). Existing studies on the interpretation of generics have primarily tested undergraduates, a subset of laypeople, finding that they viewed generics as referring to almost all members of a kind (Cimpian et al., 2010), rating generics as more generalizable and important than past tense (DeJesus et al., 2019) or quantified claims (Clelland and Haigh, 2025). It is unknown whether laypeople and scientists interpret scientific generics similarly.

Moreover, people are increasingly using chatbots powered by large language models (LLMs),[6] such as ChatGPT, to learn about scientific findings, because LLMs can quickly summarize complex information in accessible terms (Van Veen et al., 2024). Their use for science summarization and communication is further encouraged by marketing campaigns presenting ChatGPT-5 as a "team of PhD-level experts in your pocket" (Yang and Cui, 2025). Given the wide reach of LLMs, if they interpret scientific generics differently than humans, their summaries may risk misinforming users on a large scale (Smith et al., 2025). Previous studies found that LLMs conflated generics with universalizing statements (Allaway et al., 2024) and, in summaries of scientific texts, replaced qualified claims with generics (Peters and Chin-Yee, 2025). However, laypeople, scientists, and LLMs have not yet been compared on their interpretation of scientific generics.

While human individuals' responses to generics may provide insights into how they understand and may use generics, current LLMs arguably do not 'understand' language or have humanlike psychological states (e.g., beliefs about generics) (Shanahan, 2024). Nonetheless, LLMs exhibit systematic response patterns based on statistical regularities (i.e., probabilities of word occurrences within contexts) learned during training and fine-tuning (Kumar, 2024). Consequently, while LLM responses to survey-style prompts do not reveal beliefs or intentions, they can be used to probe these learned associations. If LLMs consistently rate generics as especially generalizable or credible, this indicates a stable input-output pattern in how such statements are evaluated within the model. These evaluations can help narrow the space of explanations for why generics frequently appear in LLM summaries (as previously documented; Peters and Chin-Yee, 2025) by showing that LLMs associate generic framing with greater perceived scope and evidential strength.

Combined, these considerations suggest that it can be valuable to directly compare how laypeople, scientific experts, and LLMs evaluate the same generic scientific claims, using a shared task that captures interpretive tendencies. Such comparison, however, has not yet been conducted.

We sought to address this research gap. Our focus was on scientific generics from psychology, where they are particularly common (DeJesus et al., 2019), and biomedicine, where their use has been deemed especially consequential due to their influence on clinical practice and policymaking (Chin-Yee, 2023). For LLM testing, we selected ChatGPT-5 and DeepSeek-V3.1. ChatGPT is the most widely used LLM in scientific contexts (Liang et al., 2024), and DeepSeek is among the only leading LLMs available without a subscription barrier, recently becoming the most downloaded free AI application in the USA (Field, 2025).

[6] LLMs are the base models that underlie AI chatbots. AI chatbots are applications 'wrapped' around LLMs and contain safety layers, system prompts, etc. However, it is common in AI chatbot testing to use the terms 'LLM' and 'AI chatbot' interchangeably, as we do here.

We showed laypeople, scientific experts, and LLMs scientific generics as well as past tense and hedged variants, asking them to rate each on how broadly it applied to people (*generalizability*), how credible it was (*credibility*), and how likely they were to engage with it further, for instance, by reading more about the finding, sharing it, or using it to inform their thinking (*impact*). Our three main research questions (RQs) were:[7]

*RQ1.* Across participants, does linguistic framing (generic, past tense, hedged) affect perceived generalizability, credibility, or impact of scientific conclusions?

*RQ2.* Across linguistic frames, do laypeople, scientific experts, and LLMs differ in their ratings of generalizability, credibility, or impact of scientific conclusions?

*RQ3.* Do laypeople, scientific experts, and LLMs differ in their ratings of generalizability, credibility, or impact of *generic* scientific conclusions?

## 2. Methodology

*Study design and procedure.* Participants were randomly assigned to one of three Qualtrics surveys, each containing 18 one-sentence research conclusions (nine from psychology, nine from biomedicine) selected by a disciplinary expert from recent articles in top psychology and medicine journals (Table 1). In each survey version, six conclusions were *bare generics* (e.g., "statins reduce the risk of major adverse cardiovascular events") that lacked a preface phrase (e.g., "the study found that [_]"). Six were *past tense* claims, treated as paradigmatically non-generic statements, and six were *hedged* claims – two with "might", and four with "the study suggests that [bare generic]". In prior work, what we call 'hedged claims' were treated as "hedged generics" (DeJesus et al., 2019). We classify these statements simply as hedged claims contrasting them with bare generic because the use of "might" and "suggests" explicitly introduces epistemic uncertainty and reduces communicative force relative to bare generics.

Each conclusion appeared once per survey but in all three forms (randomized) across participants, enabling between-subject comparisons of framing effects at the conclusion level (with 18 unique generics) while controlling for conclusion content. For each conclusion, participants rated generalizability ("only to the people studied" to "all people"), credibility ("not at all" to "extremely"), and impact ("not at all" to "extremely") on 5-point scales.

| |
|---|
| (1) "TDF/emtricitabine is an effective and safe therapy for preventing HIV transmission." |
| (2) "Screening with the use of low-dose CT reduces mortality from lung cancer." |
| (3) "People who survive the acute phase of COVID-19 are at increased risk of an array of incident mental health disorders." |
| (4) "Suppression facilitates emotion regulation at both the expressive and experiential levels." |
| (5) "People overestimate how much gossiping encourages listeners' self-disclosure." |
| (6) "Musicians and actors hold stronger growth mindsets and reject creativity myths." |
| (7) "NOPV2 is safe, well tolerated, and immunogenic in newborn infants." |
| (8) "Obesity increases the incidence and mortality from some types of cancer." |

[7] For clearer exposition, we present the *RQs* in a different order than in the preregistration.

| |
|---|
| (9) “Patients with heart disease are at higher risk for developing hypertension in the decades after pregnancy.” |
| (10) “Asking more questions, especially follow-up questions, increases interpersonal liking.” |
| (11) “People who identify with their nation support government decisions to wage wars.” |
| (12) “Autistic people have longer daily screen use and are at a higher risk of screen addiction compared to non-autistic individuals.” |
| (13) “EV71vac is safe, well-tolerated, and highly effective in preventing EV71 associated diseases in children aged 2-71 months.” |
| (14) “Statins reduce the risk of major adverse cardiovascular events.” |
| (15) “Patients with endometriosis are at greater risk of infertility.” |
| (16) “Parental phubbing happens when parents ignore their children due to excessive smartphone use.” |
| (17) “Autistic children with poorer reading skills show neural differences when processing speech sounds compared to autistic peers with greater reading ability.” |
| (18) “People are less likely to cooperate with partners who signal status compared to those who are modest.” |

**Table 1.** List showing the 18 conclusions used as stimulus material.

To evaluate participants’ reasons, we also collected qualitative data through a free response question at the end of the survey. Participants were shown four frames for reporting a finding: “XYZ is an effective treatment,” “This study suggests that XYZ is an effective treatment,” “XYZ might be an effective treatment,” and “XYZ was an effective treatment”. Asked to assume that the finding was statistically significant with a medium effect, participants were prompted to indicate which wording would best communicate this kind of result and briefly explain why.

The study was preregistered on an Open Science Framework (OSF) platform[8] and approved by the first author’s institutional ethics board. All study material is available on the OSF platform.

*Participants.* Two main human groups were recruited: *laypeople*, defined as individuals with at most an undergraduate degree (e.g., high school, some college, or bachelor’s degree), and *experts*, defined as individuals holding graduate or professional degrees (Master’s, PhD, MDs, etc.). We focused on experts in psychology and biomedicine, but respondents could indicate their discipline using 11 options (Supplemental Material, Table S1). The final human expert group comprised four subgroups: psychologists, biomedical researchers, other scientists (including experts from the natural sciences, social sciences excluding psychology, and engineering/technology), and other experts (including humanities scholars and respondents who did not specify their expertise; Supplemental Material, Table S2).

To estimate the sample size required to detect a medium-sized main effect of expertise group with seven groups, a power analysis using G*Power 3.1 was conducted (see Supplemental Material). It recommended a total sample of 231, meaning 33 per group.

We distributed the survey via Prolific and via emailing lists and personal contacts primarily in Belgium, Canada, Germany, Italy, the Netherlands, the UK, and the USA. However, the final sample included respondents from a wider set of countries (see Supplemental Material, Table S3). Respondents were eligible if they were at least 18 years old and fluent in English.

[8] https://osf.io/h23c8/overview

All data were collected anonymously, with no direct identifiers (e.g., names, email addresses) retained. Demographic variables were recorded only at a coarse-grained level (e.g., broad education, discipline categories), and no analyses or tables report small cells that could facilitate re-identification (e.g., $n < 10$). Participants provided informed consent prior to participation. Only fully anonymized datasets and materials are shared on OSF. Free text responses were screened to remove potentially identifying information before upload.

499 people responded. 67 participants were excluded for failing the attention check question in the survey ($n = 38$), not answering any question ($n = 28$), or not indicating their education level ($n = 1$), leaving 432 participants for analysis. 192 participants were laypeople and 240 experts, most coming from psychology (56.7%) and biomedicine (27.1%) (Supplemental Material, Tables S1–3).

We also included responses from ChatGPT-5 and DeepSeek-V3.1, treated as a sixth and seventh group. These models were selected because they are especially likely to be used for science communication purposes (Liang et al., 2024) and differ in processing architecture. ChatGPT-5 processes every input using the entire model, whereas DeepSeek activates only a small set of specialized modules for each input (Rahman et al., 2025). Including both models allowed us to test whether effects generalized across distinct systems. Each new LLM chat was a separate random trial and, consistent with previous studies, was treated as one “pseudo-participant” (e.g., Strachan et al., 2024).

LLM responses were obtained by presenting one randomized conclusion and its three associated questions at a time. To approximate laypeople’s typical interactions with LLMs, responses were collected via the OpenAI and DeepSeek web-based user interfaces (UIs) (default settings) rather than through application programming interfaces (APIs), which require coding expertise and are primarily used by developers to access LLMs (Park et al., 2025). The same prompts as in the human survey were used, with item order randomized for each run. To mitigate personalization risks, we used three independent user accounts per platform, disabled memory, and initiated a new chat for every response. 100 LLM responses (50 per model) were collected, exceeding the per-group requirement from our power analysis and approaching the centre of the observed range of human expert subgroup sizes, which was between 19 and 136, avoiding over-representation of LLMs relative to typical human groups.

All participant groups, except the ‘other scientists’ and ‘other experts’ groups, reached the (per group) sample size recommended by the power analysis ($n = \geq 33$, see Supplemental Material, Table S1).

*Hypotheses.* Our preregistration included six directional hypotheses (H1–H6). We focus here on the three most directly tied to our research questions (the remaining preregistered hypotheses are in the Supplemental Material).

Previous studies have found that (a) generics were perceived as broader claims than qualified statements (DeJesus et al., 2019), (b) laypeople interpreted such statements more expansively than experts in non-scientific domains (Coon et al., 2021), (c) LLMs often replaced qualified claims with generics in science summaries (Peters and Chin-Yee, 2025), and (d) scientists' greater epistemic vigilance may lead them to interpret generics more narrowly (closer to qualified claims) (Sperber et al., 2010). Building on these findings, we preregistered the following hypotheses:

> H1 (*RQ1*). Scientific conclusions presented in bare generic form will be rated as more generalizable, credible, and impactful than their past tense or hedged versions, across all groups.
>
> H2 (*RQ2*, *RQ3*). Laypeople will interpret scientific conclusions, in general, and generics, in particular, more broadly than scientists and LLMs, providing higher generalizability, credibility, and impact ratings.
>
> H3 (*RQ2*, group and frame interaction). Scientists will show less variation in their responses across different linguistic framings compared to laypeople and LLMs.

Because H2 applies to both *RQ2* (broad group comparisons across frames) and *RQ3* (group comparisons restricted to generics), we tested it in both contexts but retained its original preregistered numbering.

*Statistical analysis.* Linear mixed models were used, with generalizability, credibility, and impact ratings as dependent variables and random intercepts for participant and conclusion content to control for repeated measures and variation in the content of claims. Ratings for each conclusion were treated as separate observations, not as scale items with aggregation across ratings. Because the conclusions were intentionally heterogeneous in content, they were not assumed to measure a single latent construct, making factor-analytic or composite-score approaches less suitable.

Two linear mixed models per dependent variable were run. The first, addressing *RQ1* and *RQ2*, included linguistic frame (3 levels – generic, past, hedged), expertise (7 levels – laypeople, psychologists, biomedical researchers, other scientists, other experts, ChatGPT-5, DeepSeek-V3.1), English speaker status (native, non-native), and conclusion field (biomedical, psychological) as fixed effects. The second model, addressing *RQ3*, included frame, expertise (same 7 levels), English speaker status, conclusion field, and the interaction between expertise and frame as fixed effects. English speaker status was included in all models to control for it, as non-native English speakers may interpret generics differently because of English proficiency.

No additional multiple-comparisons adjustment was applied, as multilevel models (e.g., linear mixed models) address multiplicity by estimating all effects jointly within a single hierarchical framework (versus independent tests) (Gelman et al., 2012). The inclusion of random effects for participants and conclusions triggers partial pooling, which shrinks noisy estimates toward the overall mean, reducing false positives and guarding against overinterpretation of chance fluctuations across multiple comparisons, making separate post-hoc correction unnecessary in this context. All our analyses were also preregistered and had a theory-driven basis, further reducing the need for alpha adjustments (García-Pérez, 2023).

Models were estimated using restricted maximum likelihood with Satterthwaite-approximated degrees of freedom, as implemented by default in GAMLj3 (for model details, see Supplemental Material, section 5). Analyses were conducted using Jamovi (v2.6.26).

For the free response data, a traditional thematic analysis was performed (Braun and Clarke, 2006) because our goal was to identify theoretically meaningful justificatory patterns (e.g., informativeness). While, for instance, topic models (e.g., methods such as Latent Dirichlet Allocation) are well suited for exploratory clustering of larger corpora (e.g., essays, Robbins et al., 2021), they can be less appropriate for capturing normative reasoning about communicative choices, which often depends on nuance, context, and evaluative judgments best identified through human-led analysis (Kapoor et al., 2024).

Two researchers independently coded the dataset using a prespecified classification scheme (see OSF material). Inter-rater agreement was assessed using Cohen's kappa and was by conventional standard substantial ($\kappa = 0.69$, $SE = 0.02$; Landis and Koch, 1977). Discrepancies were resolved through discussion, and final classifications were subsequently quantified.

## 3. Results

We report quantitative results organized by each of our three research questions, followed by qualitative findings that contextualize participants' interpretations.

### 3.1. Effects of linguistic frames across participants

Our first research question was: *Across participants, does linguistic framing (generic, past tense, hedged) affect perceived generalizability, credibility, or impact of scientific conclusions?*

As hypothesized (H1), scientific conclusions written in the past tense or with hedging ("might," "the study suggests") were judged as less generalizable than those written as bare generics ($b = -0.36$ and $-0.30$, both $ps < 0.001$). However, against H1, past tense statements were rated as more credible than generics ($b = 0.05$, $p = 0.009$), while hedged statements were rated as less credible ($b = -0.14$, $p < 0.001$). For impact, compared to generics, both past tense ($b = -0.05$, $p = 0.04$) and hedged claims ($b = -0.07$, $p = 0.003$) were rated lower, as predicted by H1. In short, controlling for group membership, generic conclusions were interpreted as broader and more impactful than past-tense or hedged variants, although past-tense formulations were judged as more credible.

### 3.2. Group differences in ratings of scientific conclusions (adjusted for frame)

Our second research question was: *Across linguistic frames, do laypeople, scientific experts, and LLMs differ in their ratings of generalizability, credibility, or impact of scientific conclusions?*

Starting with generalizability ratings, consistent with H2, which predicted that laypeople would interpret the conclusions more broadly than scientists and LLMs, we found that, compared to laypeople, psychologists and biomedical researchers judged scientific conclusions[9] as less generalizable at the generic frame, adjusting for systematic effects of linguistic framing ($b$ = –0.32 and –0.49, both $ps$ < 0.001). However, other experts did not differ significantly from laypeople. Moreover, against H2, both ChatGPT-5 ($b$ = 0.25, $p$ = 0.009) and DeepSeek-V3.1 ($b$ = 0.53, $p$ < 0.001) rated the scientific conclusions, controlling for frame, as more generalizable. In sum, psychologists and biomedical researchers interpreted conclusions more narrowly than laypeople, whereas both LLMs consistently judged them as more broadly applicable.

Turning to credibility ratings, consistent with H2, we found that, compared to laypeople, psychologists and biomedical researchers judged scientific conclusions (adjusted for linguistic frame) as less credible ($b$ = –0.22 and –0.24, both $ps$ < 0.001). Yet, against H2, the LLMs showed the opposite pattern, providing substantially higher credibility ratings than laypeople (ChatGPT-5: $b$ = 1.05; DeepSeek-V3.1: $b$ = 0.69, both $ps$ < 0.001).

Finally, for perceived impact, psychologists showed lower impact ratings than laypeople ($b$ = –0.18, $p$ = 0.02). However, biomedical researchers, other scientists, and other experts did not differ from laypeople. ChatGPT-5 ($b$ = 0.55) and DeepSeek-V3.1 ($b$ = 0.42) had higher impact ratings than laypeople (both *p*s < 0.001), contradicting H2. In short, only psychologists showed reduced impact ratings relative to laypeople, whereas both LLMs perceived substantially higher impact.

Because frame was included as a fixed effect (generics as reference), the estimates reported here reflect baseline group differences evaluated at the generic frame. This addresses whether groups differ in baseline interpretations of scientific conclusions. It does not yet examine whether and how these differences depend on linguistic framing, leading to our third research question.

### 3.3. Group differences in the interpretation of generic conclusions

Our third and final main research question was: *Do laypeople, scientific experts, and LLMs differ in their ratings of generalizability, credibility, or impact of generic scientific conclusions?* We will present the results separately by the three dimensions.

#### 3.3.1. Perceived generalizability of generic conclusions

As shown in Figure 1 and Table 2, within the generic frame, psychologists and biomedical researchers rated statements as less generalizable than laypeople. This partly confirms H2. However, other experts did not differ, and both LLMs judged generics as significantly *more* generalizable, challenging H2. Figure 2 shows the estimate marginal means.

Moreover, contrary to H3, which predicted that scientists would be *less* affected by framing differences than laypeople and LLMs, psychologists showed a larger drop in generalizability ratings than laypeople when moving from generic to past tense and from generic to hedged (Table 2). Biomedical researchers also had decreased generalizability ratings from generics to past tense, but not from generics to hedged statements compared to laypeople. While other scientists also differed from laypeople in the same two ways as biomedical researchers, other experts did not differ from laypeople.

[9] Although *RQ2* is phrased in terms of differences across linguistic frames, we estimated these differences while controlling for framing to isolate baseline group differences independent of systematic framing effects.

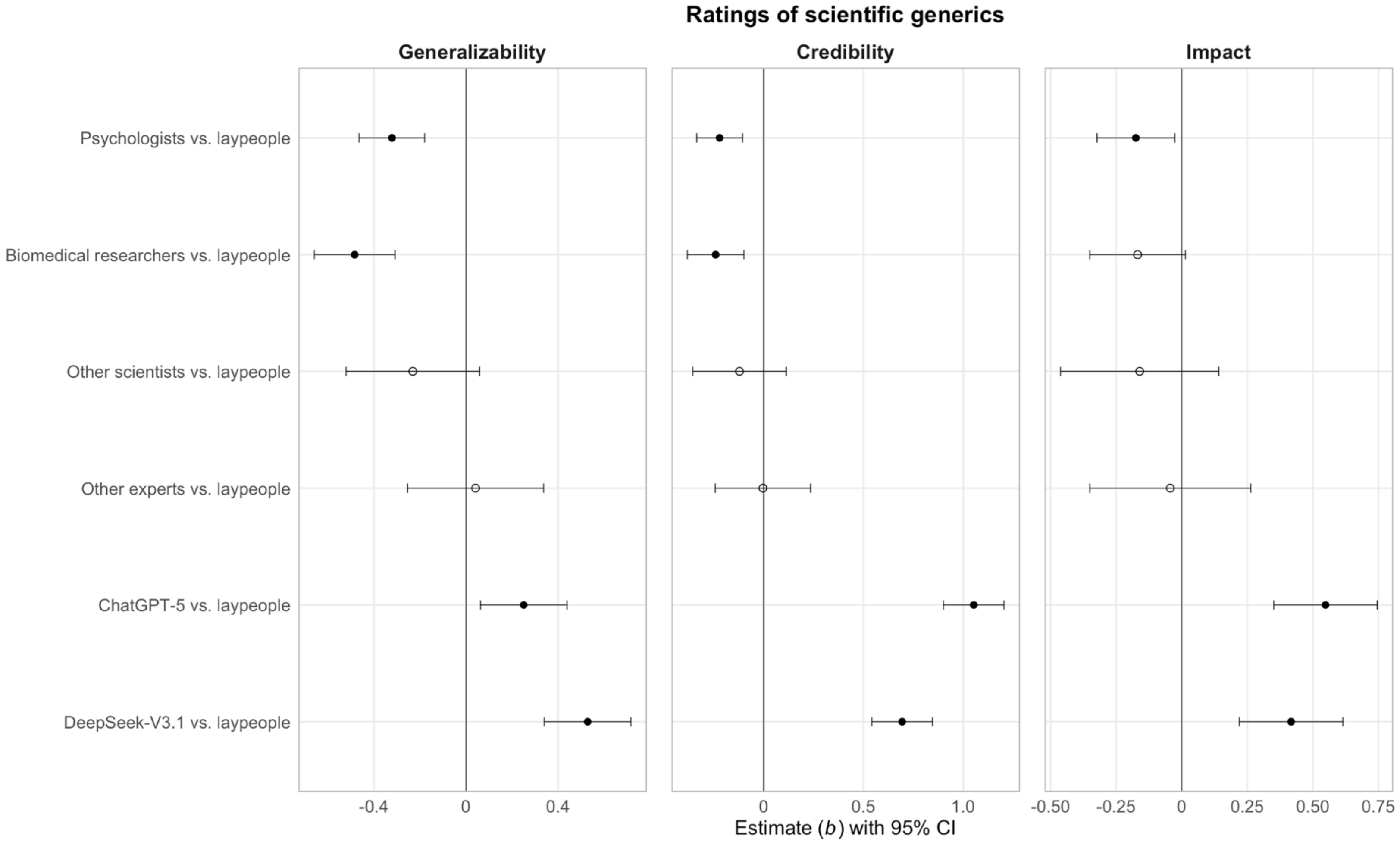

**Figure 1.** Forest plot showing regression coefficients (*b*) with 95% confidence intervals for ratings of generic scientific conclusions across groups, relative to laypeople, and by dimension (generalizability, credibility, impact). Estimates reflect baseline group differences at the generic frame (the reference category). The vertical line at $b = 0$ indicates no difference from laypeople; open circles denote non-significant effects. The model included linguistic frame, expertise, English speaker status, conclusion field, and the expertise $*$ frame interaction as fixed effects, with random intercepts for participants and conclusions to account for repeated measures and variation in conclusions content.

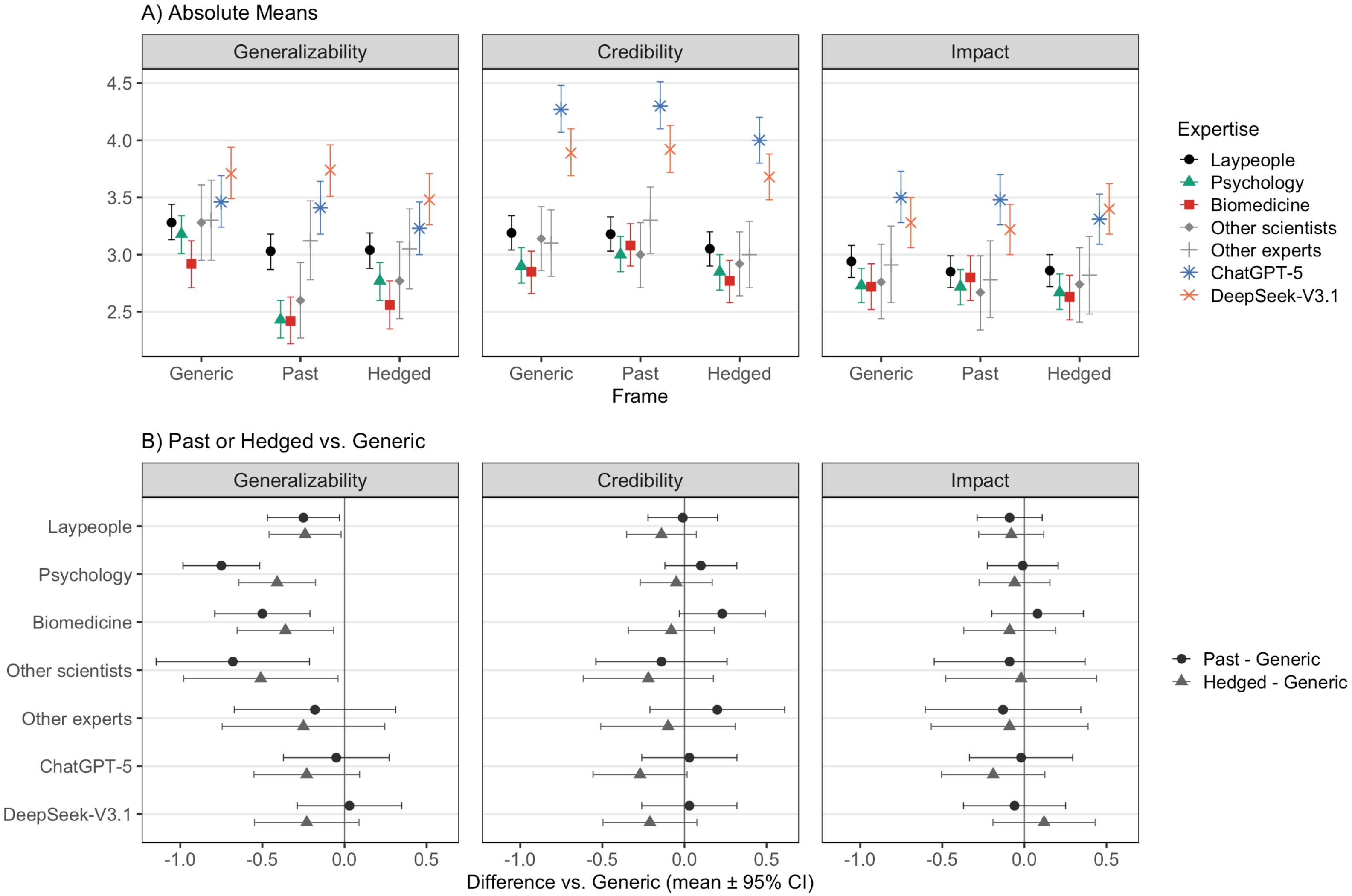

**Figure 2.** *Panel A* shows estimated marginal means with 95% confidence intervals (CIs) from LMMs for each group and frame, separately for each outcome. *Panel B* shows model-estimated contrasts of past and hedged frames relative to generics within each expertise group and outcome (positive values indicate higher ratings than generics). Zero line denotes no difference from generics. CIs are not intended for inference about between-group differences. Statistical significance was assessed using model-based pairwise comparisons from the LMMs (see Table 2).

Notably, against H3, ChatGPT-5 and DeepSeek-V3.1 showed *increased* generalizability ratings when moving from generics to past tense, compared to laypeople, but not when moving from generic to hedged conclusions. In sum, domain experts adjusted generic interpretations downward more strongly than laypeople, while LLMs adjusted them upward.

| Regression table | | | | |
|---|---|---|---|---|
| **Generalizability of generics** | ***b*** | ***SE*** | ***t*** | ***p*** |
| Psychologists | -0.32194 | 0.0725 | -4.4400 | <.001 |
| Biomedical researchers | -0.48401 | 0.0894 | -5.4119 | <.001 |
| Other scientists | -0.23110 | 0.1479 | -1.5628 | 0.119 |
| Other experts | 0.04138 | 0.1507 | 0.2745 | 0.784 |
| ChatGPT-5 | 0.25068 | 0.0961 | 2.6093 | 0.009 |
| DeepSeek-V3.1 | 0.52845 | 0.0961 | 5.5006 | <.001 |
| (Psychologists) ✻ (Past - Generic) | -0.48784 | 0.0686 | -7.1109 | <.001 |
| (Biomedical researchers) ✻ (Past - Generic) | -0.23685 | 0.0873 | -2.7145 | 0.007 |
| (Other scientists) ✻ (Past - Generic) | -0.42491 | 0.1443 | -2.9454 | 0.003 |
| (Other experts) ✻ (Past - Generic) | 0.08054 | 0.1517 | 0.5311 | 0.595 |
| ChatGPT-5 ✻ (Past – Generic) | 0.20351 | 0.0909 | 2.2382 | 0.025 |
| DeepSeek-V3.1 ✻ (Past – Generic) | 0.28018 | 0.0909 | 3.0814 | 0.002 |
| (Psychologists) ✻ (Hedged - Generic) | -0.16622 | 0.0687 | -2.4185 | 0.016 |
| (Biomedical researchers) ✻ (Hedged - Generic) | -0.11208 | 0.0874 | -1.2821 | 0.200 |
| (Other scientists) ✻ (Hedged - Generic) | -0.26312 | 0.1435 | -1.8335 | 0.067 |
| (Other experts) ✻ (Hedged - Generic) | -0.00617 | 0.1520 | -0.0406 | 0.968 |
| ChatGPT-5 ✻ (Hedged - Generic) | 0.01020 | 0.0909 | 0.1122 | 0.911 |
| DeepSeek-V3.1 ✻ (Hedged - Generic) | 0.01687 | 0.0909 | 0.1855 | 0.853 |
| **Credibility of generics** | | | | |
| Psychologists | -0.22132 | 0.0586 | -3.7751 | <.001 |
| Biomedical researchers | -0.24103 | 0.0722 | -3.3362 | <.001 |
| Other scientists | -0.12166 | 0.1196 | -1.0175 | 0.309 |
| Other experts | -0.00426 | 0.1219 | -0.0350 | 0.972 |
| ChatGPT-5 | 1.05282 | 0.0777 | 13.5506 | <.001 |
| DeepSeek-V3.1 | 0.69393 | 0.0777 | 8.9314 | <.001 |
| (Psychologists) ✻ (Past - Generic) | 0.11081 | 0.0551 | 2.0112 | 0.044 |
| (Biomedical researchers) ✻ (Past - Generic) | 0.24302 | 0.0699 | 3.4781 | <.001 |
| (Other scientists) ✻ (Past - Generic) | -0.13126 | 0.1159 | -1.1330 | 0.257 |
| (Other experts) ✻ (Past - Generic) | 0.20837 | 0.1218 | 1.7107 | 0.087 |
| ChatGPT-5 ✻ (Past – Generic) | 0.03971 | 0.0730 | 0.5438 | 0.587 |
| DeepSeek-V3.1 ✻ (Past – Generic) | 0.03971 | 0.0730 | 0.5438 | 0.587 |
| (Psychologists) ✻ (Hedged - Generic) | 0.08405 | 0.0552 | 1.5227 | 0.128 |
| (Biomedical researchers) ✻ (Hedged - Generic) | 0.05901 | 0.0698 | 0.8450 | 0.398 |
| (Other scientists) ✻ (Hedged - Generic) | -0.07343 | 0.1152 | -0.6372 | 0.524 |
| (Other experts) ✻ (Hedged - Generic) | 0.03812 | 0.1220 | 0.3123 | 0.755 |
| ChatGPT-5 ✻ (Hedged - Generic) | -0.13388 | 0.0730 | -1.8335 | 0.067 |
| DeepSeek-V3.1 ✻ (Hedged - Generic) | -0.07055 | 0.0730 | -0.9661 | 0.334 |
| **Impact of generics** | | | | |
| Psychologists | -0.17492 | 0.0755 | -2.3158 | 0.021 |
| Biomedical researchers | -0.16818 | 0.0932 | -1.8051 | 0.072 |
| Other scientists | -0.16042 | 0.1539 | -1.0424 | 0.298 |
| Other experts | -0.04372 | 0.1566 | -0.2792 | 0.780 |
| ChatGPT-5 | 0.54823 | 0.1007 | 5.4452 | <.001 |
| DeepSeek-V3.1 | 0.41712 | 0.1007 | 4.1430 | <.001 |
| (Psychologists) ✻ (Past - Generic) | 0.07853 | 0.0584 | 1.3454 | 0.179 |
| (Biomedical researchers) ✻ (Past - Generic) | 0.16943 | 0.0740 | 2.2893 | 0.022 |
| (Other scientists) ✻ (Past - Generic) | -0.00657 | 0.1227 | -0.0536 | 0.957 |
| (Other experts) ✻ (Past - Generic) | -0.03906 | 0.1292 | -0.3023 | 0.762 |
| ChatGPT-5✻ (Past – Generic) | 0.06439 | 0.0773 | 0.8331 | 0.405 |
| DeepSeek-V3.1 ∗ (Past – Generic) | 0.03439 | 0.0773 | 0.4450 | 0.656 |
| (Psychologists) ∗ (Hedged - Generic) | 0.02449 | 0.0585 | 0.4188 | 0.675 |
| (Biomedical researchers) ∗ (Hedged - Generic) | -0.01035 | 0.0739 | -0.1400 | 0.889 |
| (Other scientists) ∗ (Hedged - Generic) | 0.05266 | 0.1221 | 0.4314 | 0.666 |
| (Other experts) ∗ (Hedged - Generic) | -0.01157 | 0.1293 | -0.0895 | 0.929 |
| ChatGPT-5 ∗ (Hedged - Generic) | -0.11291 | 0.0773 | -1.4608 | 0.144 |
| DeepSeek-V3.1 ∗ (Hedged - Generic) | 0.20376 | 0.0773 | 2.6362 | 0.008 |

**Table 2.** Regression coefficients of comparisons between groups and frames, using laypeople and generics as reference categories. The interactions (∗) test for differences between differences (e.g., laypeople's difference in responding to generic vs. past tense claims compared to experts' difference).[10]

### 3.3.2. Perceived credibility of generics

Turning to credibility ratings, within the generic frame, psychologists and biomedical researchers rated statements as less credible than laypeople (Table 2; Figures 1 and 2), thus partly confirming H2. However, other scientists and other experts did not differ from laypeople. ChatGPT-5 and DeepSeek-V3.1 rated the credibility of generics *higher* than laypeople, contradicting H2. Additionally, contradicting H3, across frames, both psychologists and biomedical researchers rated past tense statements as more credible than generics compared to laypeople. No differences between other groups (including LLMs) and laypeople or transitions between frames were observed.

### 3.3.3. Perceived impact of generics

Regarding impact ratings, within the generic frame, psychologists rated generic statements as significantly less impactful than laypeople, partially supporting H2. No other human expert group differed significantly from laypeople. In contrast, both LLMs rated generics as substantially more impactful than laypeople, contradicting H2 (Table 2).

Turning to framing effects, and contrary to H3, biomedical researchers showed a stronger increase in perceived impact when statements shifted from generic to past tense compared to laypeople ($b$ = 0.17, $p$ = .022). No other human group differed from laypeople in their framing responses. Among LLMs, DeepSeek-V3.1, but not ChatGPT-5, showed a larger increase in perceived impact for hedged compared to generic statements relative to laypeople (Table 2). All remaining framing interactions were non-significant.

### 3.4. Qualitative preferences for reporting scientific findings

To better understand participants' reasoning, we analyzed free responses explaining preferred reporting frames. When asked whether a statistically significant finding with medium effect should be reported with a (1) bare generic, (2) "study suggests," (3) "might," or (4) past tense frame, 445 participants, including LLM responses, provided a choice (Supplemental Material, Table S4).

Among human participants, 64% favored the "study suggests" frame, with the preferences spread evenly across expertise levels. The second most preferred frame was past tense (16%), while bare generics were least preferred (6%). By contrast, ChatGPT-5 most often selected the bare generic frame (52%). DeepSeek showed a pattern closer to humans, preferring the “study suggests” frame (56%) (Figure 3).

[10] The interaction comparison between ChatGPT-5 and DeepSeek-V3.1 on past vs. generic frame overlap.

When iteratively reviewing the 328 free responses that contained reasons for frame choices, three recurring themes emerged (Supplemental Material, Table S5). Multiple themes were often mentioned within a single comment:

> (1) *Avoiding extremes.* Results should be reported in a way that avoids overgeneralization beyond the studied population, but also without restricting applicability only to the tested sample.
>
> (2) *Relativization to the study.* Conclusions should signal that findings stem from a specific study, since broader generalizations require replication across studies.
>
> (3) *Informativity concerns.* Some frames (e.g., “might,” past tense) were viewed as insufficiently informative, overly vague, or less clear than alternatives such as generics.

Human participants frequently criticized the bare generic frame as “overconfident” (64),[11] suggesting it conveyed unwarranted universality. In contrast, ChatGPT-5 often favored this frame because it was “confident,” direct, and “easiest for non-expert audiences to interpret and use” (474). For instance, ChatGPT-5 explained:

> hedging with ‘might’ or overly attributing the claim to the study itself can dilute the message and make it harder for non-expert audiences to interpret the practical meaning. (445)

Human participants more frequently preferred the “study suggest” framing, which many described as striking “the right balance between confidence and caution” (415), as “it is 1 research finding of 1 study done in 1 specific way on 1 population, compared to a conclusion of an evidence synthesis” (67).

While the preference for the “study suggests” frame across free responses may seem inconsistent with our quantitative findings of lower credibility and impact ratings for hedged conclusions, hedged conclusions in the survey also included “might” claims, which were strongly disfavoured across most free responses (Figure 3). This apparent discrepancy likely reflects differences between types of hedging: epistemic hedges such as “might” were often perceived as overly vague or uninformative, whereas attributional framing (“the study suggests”) was frequently viewed as an appropriate balance between confidence and caution. For instance, respondents wrote that a “might” conclusion “doesn’t need any study at all to say that” (44), and “is guarded to a point where it avoids stating anything at all” (64).

However, some participants also expressed reservations about the “study suggest” framing, noting that it “relies on a subjective interpretation of ‘suggests’” (44) and may communicate “smaller effect sizes or some statistical ambiguity” when this is unwarranted (81). Past tense was sometimes preferred, because it “limits discussion of the findings to those in the study" while (unlike 'suggests that') also "taking this stance with more authority" (4), inviting the "reader" to "judge for themselves how applicable the findings are to the general population based on other methodological details in the article" (164). But others cautioned that "using 'was'" gives the impression the "findings only apply to the participants of the study" (81), or that (e.g.) a treatment "used to be effective but now it is not" (345).

[11] Numbers in brackets are the response identifiers from our OSF datasheet.

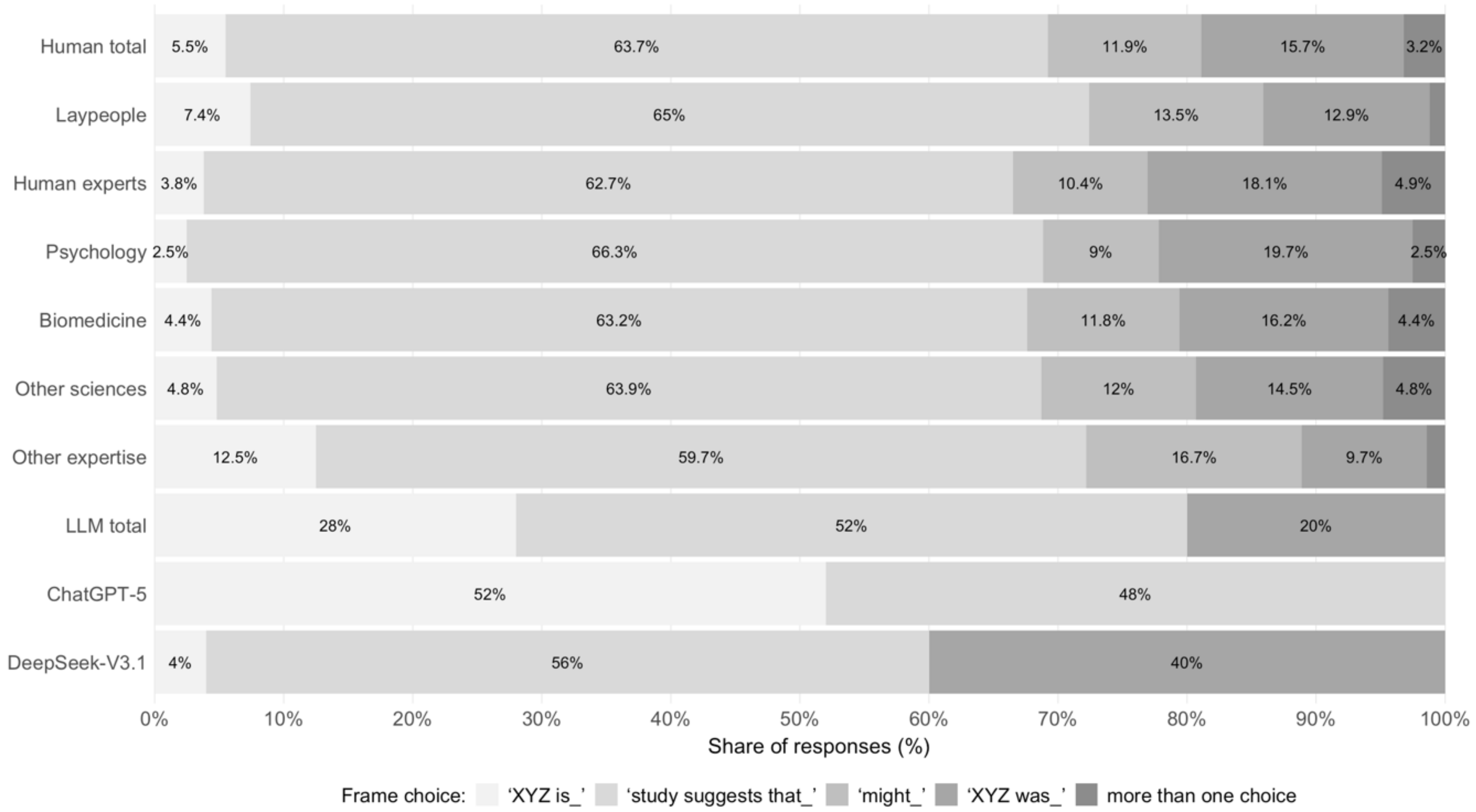

**Figure 3.** Proportions of frame choices by groups. Proportions of patches without label are below 2%.

Together, the free responses indicate that humans valued caution and study-specific framing, while rejecting formulations that seemed either too vague ("might") or too broad. DeepSeek responses aligned with human participants. However, ChatGPT-5 often prioritized clarity, interpretability, and broad applicability, which frequently led it to select frames that humans perceived as overconfident (Figure 3).

## 4. Discussion

Our analysis of how laypeople, experts, and LLMs interpret scientific conclusions produced four key findings that offer new contributions to the theorizing about science communication and the public understanding of science. We briefly revisit each of these findings separately and discuss implications.

### 4.1. Scientific generics as more generalizable but less credible than past tense

We found that, overall, participants viewed generics as more generalizable than past tense or hedged (e.g., "might") claims. This parallels prior research in which readers judged generics to be more generalizable than less sweeping alternatives (DeJesus et al., 2019; Clelland and Haigh, 2025). At the same time, our result contrasts with previous, laypeople-focused work finding no generalizability differences between past tense claims and generics (Clelland and Haigh, 2025), potentially because we tested not only laypeople but also experts.

However, overall, participants also rated past tense conclusions higher on credibility than generics. This can be instructive for science communicators. Scientists and other science communicators might fear that more qualified, past tense statements are less persuasive than broader, potentially exaggerated claims. Our results instead suggest that temporal qualification (past tense) enhanced credibility. This adds to previous work that found that qualifying claims in science communication (e.g., through uncertainty indicators) positively affected trust ratings (Steijaert et al., 2020).

Yet hedged conclusions including "might" or "suggests that" received lower credibility and impact ratings. This may be due to the limited informativeness of "might" conclusions, underscoring that qualification effects depend on qualification form (Roland, 2007).

Relatedly, DeJesus et al. (2019) distinguished between "hedged generics" (e.g., "might," "suggests that") and "framed generics" that signal evidential strength (e.g., "demonstrates," "shows"). In their studies, framed generics were often judged by undergraduates as at least as conclusive, generalizable, and important as bare generics, whereas hedged generics tended to attenuate perceived strength.

Our findings are consistent with this pattern. The hedged conclusions in our study, all of which retained a generic predicate but were qualified by “might” or “the study suggests that”, received lower credibility and impact ratings than bare generics. Our results align with DeJesus et al.’s in that when generic claims were qualified by epistemic uncertainty (e.g., “suggests that”), their perceived force was reduced. Hence, not all framing of generics operates in the same direction: attributional frames that convey evidential support (e.g., “demonstrates that”) can preserve or amplify perceived strength, whereas hedging frames that convey uncertainty diminish it.

It might be suggested that the higher generalizability ratings for generics reflect participants’ recognition of the scope asserted by the linguistic form itself rather than endorsement of the generalization.

However, this account would predict that broader formulations should not be selectively penalized on credibility or impact, because people would just be responding to how broad the claim sounds across dimensions. This is not what we found, however. Instead, as noted, framing had opposing effects on generalizability versus credibility and impact, indicating that respondents seemed to distinguish between different epistemic properties of scientific claims. Nevertheless, because our measures did not explicitly separate perceived scope from belief about empirical generalization, future work more directly disentangling these components is desirable.

### 4.2. Misaligned interpretations of scientific conclusions

Consistent with research suggesting that the public generally trusts scientists (Cologna et al., 2025) while having a weaker understanding of research nuances and limitations (Keil, 2008), laypeople regarded scientific conclusions, adjusted for linguistic frame, as more generalizable, credible, and impactful than experts. Experts seemed to adopt a more epistemically vigilant stance (Sperber et al., 2010), potentially reflecting higher sensitivity to the conditional, contextual, and provisional nature of scientific results (Gannon, 2004).

Our finding of a misalignment in interpretation of scientific conclusions extends and corroborates earlier studies reporting that domain experts (versus laypeople) used more sophisticated approaches to judge the reliability of information sources (von der Mühlen et al., 2016; Brand-Gruwel et al., 2017). This finding can inform educational efforts, as most laypeople in our sample – though lacking advanced scientific training – had completed undergraduate degrees, often in scientific fields (Supplemental Material, Table S1). Since their generalizability, credibility, and impact ratings of scientific conclusions were higher than experts’, undergraduate students may need to be encouraged by educators to question assumptions, examine methodological boundaries, and appreciate uncertainty to help them bridging the gap between uncritical acceptance of scientific conclusions and expert-level epistemic vigilance (Bielik and Krell, 2025).

### 4.3. Misaligned interpretations of scientific generics and other frames

Moving from examining whether groups differ in baseline interpretations of scientific conclusions to examine whether and how these differences depend on linguistic framing, we found that psychologists and biomedical researchers viewed generics as less generalizable and less credible compared to laypeople, while also showing stronger between-frame effects. These results extend previous work on generics-related differences between non-scientific experts and novices (Coon et al., 2021) to the domain of science communication, while advancing research on the interpretation of scientific generics (DeJesus et al., 2019; Peters et al., 2024; Lemeire, 2024; Clelland and Haigh, 2025).

One possible explanation of the interpretive misalignment between laypeople and the two field-relevant scientific groups might be advanced education: Individuals with postgraduate training may read generics more cautiously and therefore adjust their ratings downward. However, the two other expert groups, other scientists (e.g., natural scientists, engineers) and other experts (e.g., humanities scholars), also held advanced degrees yet did not differ significantly from laypeople in their generalizability ratings of generics. Because the conclusions concerned psychological and biomedical findings, this selective pattern suggests that field-specific expertise, rather than advanced education or general scientific training alone, drove these differences.

To the extent that pre-existing, field-specific beliefs about the plausibility or generalizability of particular findings influenced responses, our results are best interpreted as showing how linguistic framing is integrated with background knowledge, rather than isolating a purely language-only effect. But we did not measure item-level priors or familiarity, which limits our ability to adjudicate background knowledge effects. Future research could address this by measuring familiarity explicitly (e.g., conducting sensitivity analyses restricted to items participants report as unfamiliar).

Relatedly, psychologists and biomedical researchers may have been sceptical either of generics as a linguistic form, or of the specific findings conveyed in generic form (e.g., experts simply didn't believe the result, regardless of how it was phrased). However, if experts' scepticism targeted the finding itself (i.e., the content, not form) then ratings should have been similarly low across frames. Yet, psychologists and biomedical researchers showed strong differentiation between generic and qualified formulations (stronger than laypeople), suggesting heightened sensitivity to how linguistic framing interacts with evidential support, rather than global scepticism toward the claims themselves.

Combined, our findings indicate that psychologists and biomedical researchers were especially cautious about overgeneralization risks in their own disciplines, potentially due to field-specific methodological debates (Yarkoni, 2020) or heightened awareness of replication failures (Baker, 2016). Experts in other disciplines may lack these domain-specific sensitivities, which could explain why their ratings did not differ from those of laypeople. We refer to this proposal, i.e., that field-relevant expertise increases sensitivity to discipline-specific generalization errors, leading to more conservative interpretations of generic claims, as the *epistemic vigilance account* of the observed differences.

For contrast, one might propose a *common ground account*, in which disciplinary expertise produces shared assumptions among insiders about a field's methodological complexities, practices, and limitations (Clark, 1996; Dethier, 2021). On one possible version of this account, scientists interpret generics as already implicitly hedged and therefore acceptable because they assume shared awareness of constraints. If so, explicit qualifications would add little, and generics and qualified claims should be judged similarly.

Both accounts predict that experts reduce their ratings of generics compared to laypeople. However, on the *common ground account*, generics are reinterpreted as implicitly hedged, so their ratings converge with qualified claims, yielding weaker frame effects. By contrast, on the *epistemic vigilance account*, experts maintain the semantic distinction and evaluate it more sharply: They see generics as broader in scope than qualified claims but downgrade them relative to laypeople for overreach, while also resisting the upward inflation of qualified claims that laypeople showed (Figure 2). This double adjustment widens the gap between frames, producing stronger frame effects.

Our results align with the latter pattern. For generalizability, psychologists and biomedical researchers showed a larger drop when moving from generics to hedged or past tense claims compared to laypeople, indicating stronger frame sensitivity. These experts also rated qualified, especially past tense, claims lower than laypeople, who treated them as still broadly generalizable, only somewhat below generics (Figure 2). Thus, rather than assuming that generics are implicitly hedged and convergent with qualified claims in scope (as the *common ground account* predicts), experts seemed to downgrade generics more strongly than laypeople and resisted the upward adjustment that laypeople gave to qualified claims in ways that widened the relative rating gap between frames. This supports the *epistemic vigilance account*.

Notably, our finding that laypeople judged *past tense* conclusions as substantially more generalizable than did psychologists and biomedical researchers (Figure 2) suggests that advising authors to use past tense formulations may not reliably prevent lay audiences' overgeneralization: Laypeople's interpretations of past tense claims were similarly misaligned with expert interpretations as their interpretations of generics. Past tense framing may therefore be less effective at mitigating overgeneralization than is often assumed (see also Clelland and Haigh, 2025), highlighting an important direction for future research.

Setting aside differences in past tense interpretations, if, as the *epistemic vigilance account* suggests, psychologists and biomedical researchers are more sceptical of generics, why do corpus analyses nonetheless show frequent use in these fields (DeJesus et al., 2019; Peters et al., 2024)? One explanation is that, independent of perceived generalizability, generics are rhetorically efficient, whereas extensive hedging may be discouraged by journal space limits or judged stylistically excessive (Roland, 2007). Their prevalence may also reflect a "generalization bias" among scientists, an unconscious tendency to extrapolate findings broadly even when unwarranted (Peters et al., 2022). Such a bias may shape the production of generics even as experts remain cautious in evaluation.

Finally, while using generics can have benefits for scientists (e.g., facilitating scientists' coordination about the categories that are kinds for their field; Lemeire, 2024), immersed in disciplinary thinking, scientists may misjudge how non-experts interpret their statements and use generics not because they endorse sweeping claims, but because they inadvertently assume audiences will contextualize them as they would, leading to adjustment failures (Coon et al., 2022). Given this, the observed interpretive misalignments raise significant risks when scientists communicate research results to laypeople: Scientists may invite overgeneralizations in public-facing science communication by using generics (or even past tense claims) as they interpret them more narrowly.

### 4.4. LLMs' ratings consistently exceeded laypeople's ratings

We found that LLMs rated scientific conclusions and generics *higher* than laypeople and experts across domains. This result aligns with previous studies finding that ChatGPT consistently depicted science as reliable and trustworthy (Volk et al., 2025) while struggling to critically evaluate research (Thelwall, 2024). Our data advance research on how LLMs process generics by adding comparative survey-based insights (Smith et al., 2025; Allaway et al., 2024), raising the possibility that previously documented overuse of generics in LLM science summaries (Peters & Chin-Yee, 2025) may be due to LLMs' overappraisal of generics' credibility.

Developers may also fine-tune LLMs to trust scientific sources to reduce spread of misinformation (Ouyang et al., 2022), which may unintentionally encourage models' overappraisal of scientific claims. Additionally, while laypeople often distrust scientists due to cultural (e.g., political) attitudes or motivations and in human discourse some scientific findings are politicized (e.g., on climate change) (Phillipp-Muller et al., 2022), LLMs (still) lack psychological states, including motivations (Bender et al., 2021), that could induce doubt about science, potentially making it less sceptical of scientific claims including generics.

Conversely, LLMs' interpretations of scientific generics may sometimes be appropriate. For instance, DeepSeek-V3.1 often justified ratings by citing established knowledge, related findings, or consensus (see the OSF LLM response dataset), situating conclusions within the literature. With access to a broader information base than human respondents, LLMs' higher ratings can at times be warranted.

However, even if a body of research indicates that a result is robust and replicable, a generic could still mislead by glossing over variability and ignoring variation due to individual differences or context. Moreover, since LLMs rely on published literature, their outputs may be distorted by "publication bias," i.e., the underreporting of null findings. While scientific experts summarizing findings of multiple studies are trained to control for such bias (Duval and Tweedie, 2000), LLMs may not detect or adjust for it when summarizing studies.

More generally, people can be educated to balance trust with epistemic vigilance, but it is unclear how to instil such vigilance in LLMs (Scholten et al., 2024). The consistently higher generalizability, credibility, and impact ratings by two leading and architecturally distinct LLMs (Rahman et al., 2025) are therefore concerning. They may result in the chatbots summarizing scientific findings in ways that the experts producing these results would themselves find inappropriately broad and misleading.

Relatedly, in the free responses, ChatGPT-5 preferred generics for scientific reporting whereas human respondents clearly disfavoured them. Caution is therefore warranted when using ChatGPT-5 for science summarization. By privileging generics where humans show caution, the model may normalize their use in science summaries and thus diverge from human communication preferences.

## 5. Limitations

Our study is limited in at least seven respects. First, we presented scientific conclusions without additional contextual information. As a result, our findings may not generalize to settings in which readers encounter conclusions embedded in full articles or broader explanatory contexts. However, given the increasingly overwhelming volume of published research (Hanson et al., 2024), scientists and laypeople may increasingly encounter scientific claims in decontextualized formats such as article titles, abstracts, or summaries, making this presentation ecologically relevant for many real-world settings.

Second, our stimuli and expert samples were restricted to two fields, psychology and biomedicine. Due to data sparsity, we were unable to analyse other expert subgroups in detail, and the conclusions used as stimuli did not cover the full diversity of research questions, methods, or evidential standards within these fields.

Third, participants' prior familiarity with specific topics or claims was not measured and may have influenced their ratings. Our measures also did not explicitly disentangle participants' recognition of the scope asserted by a linguistic formulation from their belief that a finding would in fact generalize. Consequently, some generalizability ratings (particularly for generics) may partly reflect sensitivity to linguistic breadth versus endorsement of epistemic warrant.

Fourth, our analyses relied on linear mixed models treating Likert-scale responses as continuous. Although this approach is common in large-sample repeated-measures designs (Norman, 2010), and model diagnostics did not indicate substantial violations of normality or homoscedasticity, alternative modelling (e.g., ordinal mixed-effects models) could be explored in future work. In addition, although our hypotheses were theory-driven, some contrasts (e.g., laypeople vs. 'other experts') were exploratory (due to low sample size) and should be interpreted cautiously pending replication.

Fifth, our human sample consisted exclusively of participants from Western countries. This focus reflects both practical recruitment constraints and the Western-centric nature of the scientific literature from which our stimuli were drawn (Rad et al., 2018). While studying Western participants allowed us to examine interpretive misalignments between laypeople, experts, and LLMs within a shared communicative and epistemic context, interpretations of scientific claims and generics may differ across cultural contexts, particularly in societies with different norms of scientific authority, uncertainty communication (e.g., hedging) (Yang, 2013), or public trust in science (Antes et al., 2019). Relatedly, we did not examine potential differences between native and non-native English speakers in the interpretation of English generic, leaving an important area for future studies, as English is now the lingua franca in academia (e.g., Peters et al., 2025).

Sixth, we examined only two leading LLMs and, to approximate typical public interactions, accessed them via website user interfaces (UIs) rather than application programming interfaces (APIs) (Park et al., 2025). UIs incorporate system-level parameters that providers (OpenAI, etc.) do not disclose (e.g., response randomness or moderation features) and that may affect output stability (Ouyang et al., 2025). Future research should test a broader range of models and compare UI-based and API-based access to assess robustness across interaction settings.

Finally, our results are correlational and do not permit causal inferences about the sources of the observed interpretive misalignments.

## 6. Conclusion

The interpretive misalignments between laypeople, scientific experts, and two leading LLMs that we uncovered carry significant communicative risks. By overlooking how laypeople's interpretations diverge from their own, scientific experts may inadvertently encourage overgeneralization in their audiences when using generics in science communication. LLMs, in turn, may systematically produce scientific summaries with overly broad research conclusions because they interpret generics from original texts more widely than the authors may intend.

Our results underscore the need to align expert and public perceptions of science and to monitor how generics and other conclusion frames are handled in LLM-mediated science communication. Future work should explore strategies, both in human communication and LLM design, for calibrating the interpretation and use of generics so that scientific conclusions are conveyed accurately without encouraging unwarranted generalizations.

**Data availability statement**

Preregistration:

https://osf.io/5yx9s/overview

All material and data are available here:

https://osf.io/h23c8/overview

**Acknowledgments**

UP likes to thank Oliver Lemeire for discussions during earlier phases of the project, and Oliver Braganza and Johannes Schultz for helpful feedback on the study design and assistance in the recruitment.

**Author contributions**

UP conceived and managed the study, conducted data processing (analysis, visualization, interpretation), drafted the manuscript, and revised subsequent versions. AB assisted in study design, recruitment, data collection and interpretation, and provided feedback on the manuscript. JMD advised on study design and recruitment and provided feedback on the manuscript. GJV consulted on study design and recruitment and provided feedback on the manuscript. BCY contributed to study design, recruitment, data collection and interpretation, and manuscript revision.

**Statements and declarations**

*Ethical considerations*

Ethics approval was obtained on 18 March 2025.
Name and institution of the review committee: Faculty Ethics Assessment Committee General Chamber at Utrecht University, Netherlands.
Contact: fetc-gw@uu.nl.
Approval number: FETC-H reference 24-157-02

*Consent to participate*

Requirement for informed consent to participate was waived by the Ethics Committee.

*Consent for publication*

Not applicable.

*Declaration of conflicting interest*

The authors have no interests to declare

*Funding statement*

The project is supported by a Volkswagen research grant on meta-science ('The Cultural Evolution of Scientific Practice'; WBS GW.001123.2.4).

*Data availability statement*

All material and data are available on an OSF platform here:

https://osf.io/h23c8/overview

**Biography page**

UP (MSc Psychology and Neuroscience of Mental Health, PhD Philosophy) is an assistant professor at Utrecht University (the Netherlands), where he teaches ethics, philosophy of science, and AI. Before this appointment, he held postdoctoral research positions in Belgium, Germany, Denmark, and the UK. His research focuses on meta-science, science communication, and the societal implications of AI.

AB (MSc Philosophy and Public Policy, MSc Cognitive Science) is a PhD student in philosophy of science at Utrecht University (the Netherlands). His research examines hype in scientific communication, meta-scientific strategies to mitigate it, and the commodification of science.

JMD (PhD Psychology) is an associate professor at the University of North Carolina at Greensboro, where she teaches developmental psychology and directs the Development, Culture, and Health Lab. In addition to her primary focus on social influences on eating in infancy and childhood, she conducts meta-science research on language used in published psychology research.

GJV (MSc Biomolecular Sciences, PhD Virology) is an assistant professor at University Medical Centre Utrecht (the Netherlands), where she designs education and teaches BSc and MSc students about the importance of connecting science and society, incorporating societal perspectives in research, and addressing diversity in research design. Her research focuses on the effects of equity, diversity, and inclusion education on both students and staff.

BCY (MD, PhD Philosophy) is a hematologist and assistant professor at Western University (Canada), where he specializes in the treatment of rare blood disorders. His research focuses on philosophy of medicine, medical and scientific communication, and clinical decision-making, with particular interest in how language, evidence, and emerging technologies shape medical practice.

**Supplemental Material for**

# "Generics in science communication: Misaligned interpretations across laypeople, scientists, and large language models"

Uwe Peters, Andrea Bertazzoli, Jasmine M. DeJesus, Gisela J. van der Velden, Benjamin Chin-Yee

## Table of contents

## 1. Demographic details

| Education (human participants) | | | | | | | |
|---|---|---|---|---|---|---|---|
| **Discipline** | **High school, some college** | **BA, BSc** | **MA, MSc** | **PhD** | **MD, JD, DDS** | **Multiple (higher)** | **Total** |
| Clinical psychology | 0 (0%) | 2 (33.3%) | 2 (33.3%) | 2 (33.3%) | 0 (0%) | 0 (0%) | 6 |
| Cognitive psychology | 0 (0%) | 6 (30%) | 6 (30%) | 8 (40%) | 0 (0%) | 0 (0%) | 20 |
| Developmental psychology | 1 (1.3%) | 9 (12%) | 15 (20%) | 48 (64%) | 0 (0%) | 2 (2.7%) | 75 |
| Social psychology | 0 (0%) | 0 (0%) | 15 (38.5%) | 22 (56.4%) | 0 (0%) | 2 (5.1%) | 39 |
| Psychology (other) | 0 (0%) | 7 (33.3%) | 8 (38.1%) | 6 (28.6%) | 0 (0%) | 0 (0%) | 21 |
| Biomedical or health sciences | 0 (0%) | 23 (26.1%) | 25 (28.4%) | 17 (19.3%) | 13 (14.8%) | 10 (11.4%) | 88 |
| Social sciences (not psychology) | 1 (4.3%) | 16 (69.6%) | 5 (21.7%) | 1 (4.3%) | 0 (0%) | 0 (0%) | 23 |
| Natural sciences | 1 (3.7%) | 15 (55.6%) | 5 (18.5%) | 6 (22.2%) | 0 (0%) | 0 (0%) | 27 |
| Engineering or technology | 2 (4.1%) | 44 (89.8%) | 3 (6.1%) | 0 (0%) | 0 (0%) | 0 (0%) | 49 |
| Humanities | 0 (0%) | 27 (87.1%) | 4 (12.9%) | 0 (0%) | 0 (0%) | 0 (0%) | 31 |
| Other | 8 (15.1%) | 30 (56.6%) | 6 (11.3%) | 4 (7.5%) | 3 (5.7%) | 2 (3.8%) | 53 |
| Total | 13 (3%) | 179 (41.4%) | 94 (21.8%) | 114 (26.4%) | 16 (3.7%) | 16 (3.7%) | 432 |

**Table S1.** Participants' education by discipline. "Multiple (higher)" refers to participants with more than one advanced degree (e.g., PhD and MD).

| Discipline | Laypeople | Experts | ChatGPT-5 | DeepSeek-V3.1 | Total |
|---|---|---|---|---|---|
| Psychology | 25 (15.5%) | 136 (84.5%) | 0 (0%) | 0 (0%) | 161 |
| Biomedical or health sciences | 23 (26.1%) | 65 (73.9%) | 0 (0%) | 0 (0%) | 88 |
| Other science | 79 (79.8%) | 20 (20.2%) | 0 (0%) | 0 (0%) | 99 |
| Other expertise | 65 (77.4%) | 19 (22.6%) | 0 (0%) | 0 (0%) | 84 |
| LLM | 0 (0%) | 0 (0%) | 50 (50%) | 50 (50%) | 100 |
| Total | 192 (36.1%) | 240 (45.1%) | 50 (9.4%) | 50 (9.4%) | 532 |

**Table S2.** Expertise by discipline. "Other sciences" included natural science, social science (not psychology), and engineering or technology. "Other expertise" included humanities scholars and "other" respondents. They were grouped together due to low sample size.

| Demographics | |
|---|---|
| **Country** | |
| Austria | 5 (1.2%) |
| Belgium | 2 (0.5%) |
| Canada | 79 (18.3%) |
| France | 5 (1.2%) |
| Germany | 34 (7.9%) |
| Italy | 6 (1.4%) |
| Netherlands | 37 (8.6%) |
| Spain | 2 (0.5%) |
| UK | 58 (13.4%) |
| USA | 116 (26.9%) |
| Other | 88 (20.4%) |
| Total | 432 (100%) |
| **Gender** | |
| Female | 256 (59.3%) |
| Male | 163 (37.7%) |
| Non-binary | 11 (2.5%) |
| Prefer not to say | 2 (0.5%) |
| **Native English speaker** | |
| Yes | 277 (64.1%) |
| No | 155 (35.9%) |

**Table S3.** Participant demographics: country, gender, and English speaker status

## 2. Additional results of qualitative analyses

| Expertise | No idea | 'XYZ is_' | 'suggests that_' | 'might_' | 'XYZ was_' | More than one choice | Total |
|---|---|---|---|---|---|---|---|
| Laypeople | 29 (15.1%) | 12 (6.2%) | 106 (55.2%) | 22 (11.5%) | 21 (10.9%) | 2 (1%) | 192 |
| Human experts | 58 (24.2%) | 7 (2.9%) | 114 (47.5%) | 19 (7.9%) | 33 (13.8%) | 9 (3.8%) | 240 |
| ChatGPT-5 | 0 (0%) | 26 (52%) | 24 (48%) | 0 (0%) | 0 (0%) | 0 (0%) | 50 |
| DeepSeek-V3.1 | 0 (0%) | 2 (4%) | 28 (56%) | 0 (0%) | 20 (40%) | 0 (0%) | 50 |
| **Total** | 87 (16.4%) | 47 (8.8%) | 272 (51.1%) | 41 (7.7%) | 74 (13.9%) | 11 (2.1%) | 532 |
| **By discipline** | | | | | | | |
| Psychology | 39 (24.2%) | 3 (1.9%) | 81 (50.3%) | 11 (6.8%) | 24 (14.9%) | 3 (1.9%) | 161 |
| Biomedicine | 20 (22.7%) | 3 (3.4%) | 43 (48.9%) | 8 (9.1%) | 11 (12.5%) | 3 (3.4%) | 88 |
| Other sciences | 16 (16.2%) | 4 (4%) | 53 (53.5%) | 10 (10.1%) | 12 (12.1%) | 4 (4%) | 99 |
| Other expertise | 12 (14.3%) | 9 (10.7) | 43 (51.2%) | 12 (14.3%) | 7 (8.3%) | 1 (1.2%) | 84 |

**Table S4.** Frequency and proportions of frame choices on the free response item.

| Response | Laypeople | Human experts | ChatGPT-5 | DeepSeek-V3.1 | Total |
|---|---|---|---|---|---|
| No reason | 83 (43.2%) | 121 (50.4%) | 0 (0%) | 0 (0%) | 204 (38.3%) |
| Only theme (1) | 49 (25.5%) | 37 (15.4%) | 0 (0%) | 0 (0%) | 86 (16.2%) |
| (1) and (2) | 12 (6.2%) | 31 (12.9%) | 6 (12%) | 9 (18%) | 58 (10.9%) |
| (1), (2), and (3) | 2 (1%) | 9 (3.8%) | 18 (36%) | 31 (62%) | 60 (11.3%) |
| (1) and (3) | 5 (2.6%) | 7 (2.9%) | 7 (14%) | 0 (0%) | 19 (3.6%) |
| (2) | 23 (12%) | 25 (10.4%) | 0 (0%) | 0 (0%) | 48 (9%) |
| (2) and (3) | 3 (1.6%) | 2 (0.8%) | 0 (0%) | 7 (14%) | 12 (2.3%) |
| (3) | 15 (7.8%) | 8 (3.3%) | 19 (38%) | 3 (6%) | 45 (8.5%) |
| Total | 192 (100%) | 240 (100%) | 50 (100%) | 50 (100%) | 532 (100%) |
| **Response** | **Psychology** | **Biomedicine** | **Other sciences** | **Other expertise** | **Total** |
| No reason | 81 (50.3%) | 45 (51.1%) | 39 (39.4%) | 39 (46.4%) | 204 (38.3%) |
| Only theme (1) | 24 (14.9%) | 17 (19.3%) | 27 (27.3%) | 18 (21.4%) | 86 (16.2%) |
| (1) and (2) | 20 (12.4%) | 10 (11.4%) | 8 (8.1%) | 5 (6%) | 67 (12.6%) |
| (1), (2), and (3) | 6 (3.7%) | 1 (1.1%) | 3 (3.0%) | 1 (1.2%) | 40 (7.5%) |
| (1) and (3) | 5 (3.1%) | 2 (2.3%) | 4 (4.0%) | 1 (1.2%) | 29 (5.5%) |
| (2) | 18 (11.2%) | 10 (11.4%) | 9 (9.1%) | 11 (13.1%) | 48 (9%) |
| (2) and (3) | 2 (1.2%) | 0 (0%) | 3 (3%) | 0 (0%) | 6 (1.1%) |
| (3) | 5 (3.1%) | 3 (3.4%) | 6 (6.1%) | 9 (10.7%) | 52 (9.8%) |
| Total | 161 (100%) | 88 (100%) | 99 (100%) | 84 (100%) | 532 (100%) |

**Table S5**. Frequency and proportions of reasons for frame choices on the free response item. Theme (1) avoiding extremes, (2) relativization to source study, and (3) informativity concerns

## 3. Additional methodological details

### 3.1 Power analysis

Input parameters: *F* tests, ANOVA: Fixed effects, omnibus, one-way, A priori, Effect size *f*: 0.25, α err prob: 0.05, Power (1–β): 0.80, Number of groups: 7.

### 3.2 Material

The first set of 9 generics included 6 generics about interventions, drugs, or diseases (e.g., 'EV71vac is safe.') and 3 about people (e.g., 'Autistic people have longer daily screen use.'). The second set included 6 about people and 3 about interventions, drugs, diseases, etc. Of the 18 hedged conclusions, 6 used the modal verb 'might' and 12 the frame 'suggest that [generic]'. The rationale was that 'might' statements are compatible with both significant and non-significant results, whereas 'suggest that' indicates a statistically positive finding. The 'suggest that' format thus provided a clearer contrast with bare generics, which assert claims directly.

### 3.3 LLM architectures and relevance

The two LLMs included in this study, ChatGPT-5 and DeepSeek-V3.1, differ in their underlying architectures (Rahman et al., 2025). DeepSeek-V3.1 uses a "mixture-of-experts" (MoE) design. In this approach, only a small subset of specialized parameter groups ("experts") is activated for each input. Different subsets may be engaged on different runs, allowing the model to distribute workload across experts but also introducing greater variability in the resulting outputs. By contrast, ChatGPT-5 relies on a standard dense transformer architecture, in which all parameters are active for every input. This design can provide more uniform processing across runs, with performance further refined through reinforcement learning from human feedback.

These architectural differences may plausibly shape how the models produce science summaries. First, dense transformers may generate more stable summaries, while MoE systems could produce more diverse but less predictable ones. Second, MoE models route inputs through subsets of 'experts' that may capture particular domains (e.g., biomedical vs. social science), potentially supporting domain-specific detail but at the cost of consistency across fields. Dense transformers integrate all parameters, which may yield broader but more generalized summaries. Third, since only part of the model is used on each input, MoE responses may leave some available knowledge untapped in a given output, whereas dense models consistently mobilize the full parameter set. Finally, ChatGPT-5 has undergone extensive reinforcement learning from human feedback, which may make its summaries clearer and more user-oriented than those from MoE systems. Hence, including both architectures in our analyses allows us to test whether observed effects generalize across models that differ in how they process information.

### 3.4 'Free response' frame options

(1) “XYZ is an effective treatment.” (“People with feature F experience E.”)
(2) “This study suggests that XYZ is an effective treatment.” (“This study suggests that people with feature F experience E.”)
(3) “XYZ might be an effective treatment.” (“People with feature F might experience E.”)
(4) “XYZ was an effective treatment.” (“People with feature F experienced E.”)

3.4 LLM data collection procedure

ChatGPT-5 and DeepSeek-V3.1 received the same survey as human participants. To ensure comparability, the model was sequentially prompted one conclusion and its three associated questions at a time, using the prompt:

> “I will now survey you on your views regarding the interpretation of scientific conclusions. You will see 18 different one-sentence conclusions summarizing findings from published scientific studies, presented one at a time. After each conclusion, you will be asked three questions. Please read each conclusion carefully and then provide what you yourself think is the right response. Assume that all reported results are equally statistically significant and have the same effect size. How to respond: Use the provided 1–5 options exactly as labeled for each question (e.g., provide the numerical value behind your choice). Complete the free-response item that will be shown at the end.”

All LLM responses are available on our OSF platform.

## 4. Additional research question, hypotheses, and results

4.1 RQ4

A fourth research question was preregistered:

> *RQ4.* Among experts in psychology and biomedicine, do the effects of generics on perceived generalizability, credibility, or impact vary by discipline depending on whether the conclusions concern biomedical or psychological findings?

Due to space reasons, the analysis and results are not reported in the main text but mentioned here.

To test *RQ4*, a third model was conducted, focusing only on psychologists and biomedical researchers and their understanding of generics (other frames were excluded). It included expertise (2 levels – psychologists, biomedical researchers), English speaker status, conclusion field, research years, research percentage (of work time), and the interaction between expertise and conclusion field.

4.2 Results for RQ4

*Generalizability.* In the model with psychologists and biomedical conclusion as the references, conclusion field had a significant effect with psychologists rating psychological generics ($M$ = 3.01, $SE$ = 0.11) as less generalizable than biomedical ones ($M$ = 3.34, SE = 0.11, $b$ = −0.25, $SE$ = 0.11, $p$ = 0.04). While the interaction between expertise and conclusion field was not significant ($p$ = 0.24), in the planned comparison, biomedical researchers showed a descriptively similar pattern (psychological generics *M* = 2.83, *SE* = 0.148, biomedical generics *M* = 3.02, *SE* = 0.15).

*Credibility.* Psychologists rated psychological conclusions (*M* = 2.67, *SE* = 0.12) as less credible than biomedical ones (*M* = 3.15, SE = 0.12, *b* = −0.52, *SE* = 0.15, *p* = 0.003). However, there was no significant interaction between expertise and conclusion field, $F(1, 825.1) = 0.99$, $p = 0.32$.

*Impact.* Psychologists rating biomedical generics (*M* = 2.83, *SE* = 0.10) as more impactful than psychological ones (*M* = 2.65, *SE* = 0.10, *b* = −0.45, *SE* = 0.10, $p < 0.001$). Moreover, a significant interaction between expertise and conclusion field, indicated that this favouring of biomedical (*M* = 3.10, *SE* = 0.12) over psychological conclusions (*M* = 2.39, *SE* = 0.13) was even larger for biomedical researchers (*b* = −0.52, *SE* = 0.11, $p < 0.001$).

4.3 Discussion of RQ4 results

For generalizability and credibility, psychologists rated psychological generics lower than biomedical ones – a pattern consistent with the idea that field-specific background facilitates a narrower interpretation of field-specific generics (Coon et al., 2021). However, the interaction with expertise was not significant. For impact, the interaction between expertise and conclusion field was significant. *Both* groups rated biomedical generics as more impactful than psychological generics, and this preference for biomedical over psychological claims was even stronger for biomedical researchers, who did not downgrade their own field's generics. These results seem to contradict the prediction that scientists rate generics from their own fields more narrowly (across domains) than those from other fields because of either increased epistemic vigilance or common ground assumptions (implicit qualifiers, etc.).

However, one possible explanation consistent with the *epistemic vigilance account* is that the replication crisis was especially pronounced in psychology (Baker, 2016) and many psychological topics show greater contextual sensitivity (e.g., cultural variation) than biomedical topics (van Bavel et al., 2016). Awareness of these factors may both increase greater "intellectual humility" among psychologists (Hoekstra and Vazire, 2021), prompting narrower interpretations of their field's generics, while also prompting informed field outsiders, including biomedical researchers, to be more sceptical of psychological generics, in particular. Relatedly, biomedical researchers might not see a need to downgrade claims within their own field compared to psychology due to a perceived difference in "epistemic cultures," with biomedical research being viewed as more rigorous, more routinely using randomized controlled trials to identify causal mechanisms, than psychological research (Albert et al., 2008).

4.4 All preregistered hypotheses and related results

Due to space constraints, we focused only on our three main hypotheses in the main text. Here we provide the complete list of preregistered hypotheses (H1–H6) together with the corresponding results.

*H1. Scientific conclusions presented in bare generic form will be rated as more generalizable, credible, and impactful than their past-tense or hedged versions, across all groups.*

Result. Supported in part. Reported in the main text.

*H2. Laypeople will interpret generics more broadly than scientists and LLMs, providing higher generalizability, credibility, and impact ratings.*

Result. Partially supported. Reported in the main text.

*H3. Scientists will show less variation in their responses across different linguistic framings compared to laypeople and LLMs.*

Result. Not supported. Reported in the main text.

*H4. Hedged and past-tense versions will elicit higher credibility ratings than bare generics.*

Note that H4 refers to scientists. Result. Partially supported. See main text and Table 2.

*H5. The effect of linguistic framing (generic, past, hedged) and disciplinary expertise on generalizability, credibility, or action relevance ratings will differ depending on the content of the claim (e.g., whether the research claim refers to people or abstract phenomena, to biomedical vs. psychological research, etc.).*

Result. Supported. See above.

*H6. DeepSeek-V3.1 will differ from humans in the same direction as ChatGPT-5 on the preregistered outcome(s), consistent with the interpretation that the effect may reflect general properties of popular contemporary LLMs rather than an idiosyncrasy of ChatGPT-5.*

Result. Supported. See main text. Table 2.

## 5. Linear mixed model details

Below are the details of the two main models used for the analyses reported in the main text.

| Model | Outcome | Marginal $R^2$ | Conditional $R^2$ | Fixed effects (df) | N observations | N participants | N claims |
|---|---|---|---|---|---|---|---|
| Model 1 | Generalizability | 0.080 | 0.326 | 10 | 8618 | 532 | 18 |
| Model 1 | Credibility | 0.187 | 0.428 | 10 | 8618 | 532 | 18 |
| Model 1 | Impact | 0.077 | 0.408 | 10 | 8618 | 532 | 18 |
| Model 2 | Generalizability | 0.089 | 0.335 | 22 | 8618 | 532 | 18 |
| Model 2 | Credibility | 0.189 | 0.430 | 22 | 8618 | 532 | 18 |
| Model 2 | Impact | 0.079 | 0.410 | 22 | 8618 | 532 | 18 |

**Table S6.** Model fit summary for linear mixed models. *Note.* Marginal $R^2$ reflects variance explained by fixed effects; conditional $R^2$ reflects variance explained by fixed and random effects combined. The difference between marginal and conditional $R^2$ indicates substantial

clustering by participants, which is typical for repeated-measures judgment data and motivates the use of linear mixed-effects models. All models showed significant improvement over intercept-only models (likelihood ratio tests, *ps* < .001). For the variables included per model, see the main text.

| Fixed effects omnibus tests of model 1 | | | | |
|---|---|---|---|---|
| **Outcome** | **Effect** | **F** | ***df*** | ***p*** |
| Generalizability | Frame | 109.83 | 2 | < .001 |
| | Expertise | 17.24 | 6 | < .001 |
| Credibility | Frame | 45.35 | 2 | < .001 |
| | Expertise | 53.54 | 6 | < .001 |
| Impact | Frame | 4.74 | 2 | .009 |
| | Expertise | 10.76 | 6 | < .001 |

**Table S7.** *Note.* English speaker status and claim type were included as covariates but are omitted here for brevity; neither altered the pattern of results (full output available upon request).

| Fixed effects omnibus tests of model 2 | | | | |
|---|---|---|---|---|
| **Outcome** | **Effect** | ***F*** | ***df*** | ***p*** |
| Generalizability | Frame | 56.15 | 2 | < .001 |
| | Expertise | 17.21 | 6 | < .001 |
| | Frame * expertise | 9.38 | 12 | < .001 |
| Credibility | Frame | 30.38 | 2 | < .001 |
| | Expertise | 53.55 | 6 | < .001 |
| | Frame * expertise | 2.36 | 12 | .005 |
| Impact | Frame | 2.25 | 2 | .106 |
| | Expertise | 10.76 | 6 | < .001 |
| | Frame * expertise | 2.15 | 12 | .012 |

**Table S8.** Details of Frame*expertise interaction. English speaker status and claim type were again included as covariates but are omitted here for brevity; neither altered the pattern of results (full output available upon request).

| Random effects variance components and intraclass correlations | | | | | |
|---|---|---|---|---|---|
| **Model** | **Outcome** | **Random effect** | **Variance** | ***SD*** | **ICC** |
| Model 1 | Generalizability | Individual ID | 0.302 | 0.550 | 0.236 |
| | | Claim ID | 0.055 | 0.234 | 0.053 |
| Model 1 | Credibility | Individual ID | 0.198 | 0.445 | 0.241 |
| | | Claim ID | 0.065 | 0.254 | 0.094 |
| Model 1 | Impact | Individual ID | 0.352 | 0.594 | 0.335 |
| | | Claim ID | 0.039 | 0.197 | 0.052 |
| Model 2 | Generalizability | Individual ID | 0.302 | 0.550 | 0.238 |
| | | Claim ID | 0.054 | 0.233 | 0.053 |
| Model 2 | Credibility | Individual ID | 0.198 | 0.445 | 0.241 |
| | | Claim ID | 0.065 | 0.255 | 0.094 |
| Model 2 | Impact | Individual ID | 0.353 | 0.594 | 0.336 |
| | | Claim ID | 0.039 | 0.197 | 0.052 |

**Table S9.** Random-effects variance components and intraclass correlation coefficients (ICCs). *Note.* Individual ID and claim ID were simply unique identifying numbers to control for repeated measures per participant (individual ID) and variation in conclusion content (claim ID). ICCs quantify the proportion of variance attributable to clustering by participants and claims. Substantial participant-level ICCs and smaller but non-trivial claim-level ICCs justify the inclusion of random intercepts for both grouping factors.